\documentclass[aps,prb,showpacs,twocolumn]{revtex4}
\usepackage{graphicx}
\begin{document}

\newcommand{\figureheight}{8.2 cm}
\newcommand{\putfig}[2]{\begin{figure}[h]
        \special{isoscale #1.bmp, \the\hsize \figureheight}
        \vspace{\figureheight}
        \caption{#2}
        \label{fig:#1}
        \end{figure}}

\newcommand{\eqn}[1]{(\ref{#1})}

\newcommand{\be}{\begin{equation}}
\newcommand{\ee}{\end{equation}}
\newcommand{\bea}{\begin{eqnarray}}
\newcommand{\eea}{\end{eqnarray}}
\newcommand{\bean}{\begin{eqnarray*}}
\newcommand{\eean}{\end{eqnarray*}}

\newcommand{\nn}{\nonumber}




\title{Filtering of spin currents based on  ballistic ring}
\author{S. Bellucci$^a$  and P. Onorato$^{a,b}$\\}
\address{
        $^a$INFN, Laboratori Nazionali di Frascati, P.O. Box 13, 00044 Frascati, Italy. \\
        $^b$Department of Physics "A. Volta", University of Pavia, Via Bassi 6, I-27100 Pavia, Italy.
}
\date{\today}

\begin{abstract}
Quantum interference effects in rings provide suitable means for
controlling spin at mesoscopic scales. Here we apply such a control
mechanism  to the spin-dependent  transport in  a ballistic quasi
one dimensional ring patterned in two dimensional electron gases
(2DEGs). The study  is essentially based on the {\it natural}
spin-orbit (SO) interactions, one  arising from the laterally
confining electric field { ($\beta$ term) and the other due to to
the quantum-well potential  that confines electrons in the 2DEG (
conventional Rashba SO interaction or $\alpha$ term).} We focus on
single-channel transport and solve analytically the spin
polarization of the current. As an important consequence of the
presence of spin splitting, we find the occurrence of spin dependent
current oscillations.

We analyze 
the transport in the presence of one  non-magnetic obstacle in the
ring. We demonstrate that a
 spin polarized current can be induced when an unpolarized charge current
is injected in the ring, by focusing on the central role that the
presence of the obstacle plays.
\end{abstract}

\pacs{72.25.-b, 72.20.My, 73.50.Jt}

\maketitle

\section{Introduction}

In recent years both experimental and theoretical physics
communities have devoted a great deal of attention to  the field of {\it
quantum electronics} \cite{AAFKN98}.  In particular a big effort
has been devoted to the study and the realization of electric
field controlled spin based devices\cite{spintro}. The main
problem raised in this field is the generation of spin-polarized
carriers and their appropriate manipulation. In order to realize a
fully spin based circuitry, the interplay between spin-orbit (SO)
coupling and quantum confinement in semiconductor heterostructures
can provide a useful tool to manipulate the spin degree of freedom
of electrons  by coupling to their orbital motion, and vice versa.

Recently  many works have been focusing on the so called spin Hall
effect \cite{[3],[7],cul,noish} and most of the
implementations in two dimensional electron gases (2DEGs) proposed for
the spin manipulation are mainly based upon the SO interaction, which  can
be seen as the interaction of the electron spin with the magnetic
field appearing in the rest frame of the electron. The SO
Hamiltonian reads \cite{Thankappan}
\begin{equation}
\hat H_{SO} = -\frac{\lambda_0^2}{\hbar}\;e{\bf E}({\bf r})
\left[\hat{{\bf \sigma}}\times \left(\hat{\bf p}+\frac{e}{c}{\bf
A}({\bf r})\right)\right]. \label{H_SO}
\end{equation}
Here ${\bf E}({\bf r})$ is the electric field, $\hat{{\sigma}}$ are
the Pauli matrices, {$\hat{\bf p}$ is the canonical momentum
operator,  ${\bf A}({\bf r})$ is a vector potential,  ${\bf r}$ is
a 3 dimensional position vector } and $\lambda_0^2= \hbar^2/(2m_0
c)^2$, where $m_0$ denotes the
electron mass in vacuum. In materials $m_0$ and $\lambda_0$ are replaced by their
effective values $m^*$ and $\lambda$.

In this paper  { we consider low dimensional electron systems formed
by quasi-one-dimensional (Q1D) devices patterned in 2DEGs. In such
systems there can be  different types of {\it natural} SO
interaction, such as: (i) the so-called Dresselhaus term which
originates from the inversion asymmetry of the zinc-blende
structure\cite{3t}, (ii) the Rashba ($\alpha$-coupling) term due to
the quantum-well potential~\cite{Kelly} that confines electrons to a
2D layer, and (iii) the confining ($\beta$-coupling) term arising
from the in-plane electric potential that is applied to squeeze the
2DEG into a quasi-one-dimensional channel~\cite{Thornton,Kelly}.}

In this article we focus on the aspects of spin-interference in
ballistic Q1D ring geometries with two leads subject to natural
$\alpha$ and  $\beta$-SO coupling. In fact coherent ring conductors
enable one to exploit the distinct interference effects of electron
spin {\em and} charge which arise in these doubly connected
geometries. This opens up the area of spin-dependent Aharonov-Bohm
physics, including topics such as Berry phases, \cite{B84,bphase}
spin-related conductance modulation,\cite{NMT99,MSC99} persistent
currents, \cite{LGB90,SGZ03} spin filters \cite{PFR03} and
detectors,\cite{ID03} spin rotation,\cite{MSC02,CMC03}  and spin
switching mechanisms \cite{FHR01,FHR03,HSFR03}. \

In some earlier papers \cite{FRURIC} the spin-induced modulation
of unpolarized currents, as a function of the Rashba coupling
strength, was discussed, often in the presence of an external
magnetic field. In this paper  we present a different mechanism
based on the {\it natural} constant  Rashba coupling, without the
help of an external magnetic field. Here we also analyze the
effects due to the $\beta$ coupling. As it was discussed in
several papers \cite{noiq,qse,iii} the in-plane electric
potential, applied to patterned Q1D devices, can yield a high
electric field in the plane of the 2DEG, leading to a sizeable
$\beta$ term. In the above cited references, where   this SO term
was investigated by taking into account the sole confining
potential, it was demonstrated that in some devices (such as a
narrow Q1D wire) the effect of the $\beta$-SO term is analogous to
the one of a uniform effective magnetic field, $B_{eff}$,
orthogonal to the 2DEG ($x-y$ plane), and directed upward or
downward according to the spin polarization along the $z$
direction.

The goals of the following treatment are: (a) checking the
presence of the spin splitting in a Q1D ring due to the $\beta$
and (b) to the {\it natural} $\alpha$ SO coupling; (c)
investigating quantum interference effects in rings; (d) analyzing
the spin-induced modulation of unpolarized currents due to the SO
term; (e) the discussion of the transport in the presence of a
non-magnetic obstacle.

In order to pursue our aims we first analyze the $\beta$ coupling
case, and then we discuss the apparently more difficult case of
the $\alpha$ coupling.

 \

In section II we discuss the analogies between the presence of a
$\beta$-SO coupling and a transverse magnetic field in a Q1D
narrow channel. Thus, we introduce the Hamiltonian for the Q1D ring,
in order to calculate  the  eigenvalues and eigenstates and the
spin splitting. In section III we present the ballistic approach
to the transport through the ring and the quantum interference
effects by analyzing the oscillations in the transmission.
 In section IV we
discuss the possible  spin-induced modulation of unpolarized
currents also in the presence of a non-magnetic obstacle. In
section V we extend our analysis to the $\alpha$ (Rashba) coupling
by showing the  analogies with the $\beta$ case. We demonstrate
how the presence of a non-magnetic obstacle can produce a
significant spin current by giving a novel mechanism for the ring
based quantum spin filtering.

\section{$\beta$ SO coupling: model and relevant parameters}

\subsection{$\beta$-SO coupling and effective magnetic field}
In this section  we neglect the $\alpha$ (Rashba) coupling and the
Dresselhaus term, so that the SO Hamiltonian in Eq.\ref{H_SO}
results very simplified\cite{morozb}
\begin{equation}
\hat H_{SO}^\beta = \frac{\lambda^2}{\hbar} \hat{\sigma_z}
 \left[{{\nabla V_c({\bf r})}}\times\left(\hat{\bf
p}+\frac{e}{c}{\bf A} \right)\right]_z. \label{Hb}
\end{equation}
We can limit ourselves to the $z$ component, because the motion
perpendicular to the 2DEG is quantum mechanically frozen out (i.e.
with a mean value $\langle p_z\rangle =0$ in the ground state, for
the potential well in the $z$ direction), while we assume that no
external magnetic field is present so that ${\bf A}=0$. Notice
that $S_z$ commutes with the Hamiltonian in Eq.~\ref{Hb}, implying
that the $\hat{z}$ component of the spin is preserved in the
motion through the device. Thus the total Hamiltonian of an
electron moving in a confining potential $V_c(r)$  is equivalent
to that of a charged particle in a transverse magnetic field, but
here the sign of ${B}_{eff}({\bf r})$ depends on the direction of
the spin along $\hat{z}$\cite{noiq}.

\subsection{A Q1D channel}
The basic brick of our device are narrow  quantum wires (QWs),
that are devices of width $W$ less than $1000 \AA$\cite{thor} and
length up to some microns  (here we think to a QW where $W \sim
5-100 nm$). In these devices quantum effects are affecting
transport properties. In fact,  because of the confinement of
conduction electrons in the transverse direction of the wire,
their transverse energy is quantized into a series of discrete
values.  From a theoretical point of view a QW is usually defined
by a parabolic confining potential along the transverse direction
$\hat{x}$,  with force $\omega_d$\cite{me} i.e.
$V_c(x)=\frac{m^*}{2}\omega_d^2 x^2$.

 In the special case of a QW $e {\bf \nabla}
V_c({\bf r})\equiv m^* \omega_d^2(x,y,0)$ thus
\begin{equation}
{B}_{eff} = \frac{\lambda^2}{\hbar}\;\frac{{m^*}^2\omega_d^2
c}{e}\equiv\frac{\beta}{\hbar l_\omega}\frac{m^*c}{e},\label{Beff}
\end{equation}
where $l_\omega=\sqrt{\hbar/m^*\omega_d}$, while
$\beta\equiv\lambda^2 {m^*}\omega_d^2 l_\omega$. Next, we introduce
the effective cyclotron frequency $\omega_c=\frac{\beta}{\hbar
l_\omega}$ ($\omega_c/\omega_d=\lambda^2/l_\omega$), the related
frequency $\omega_0^2=\omega_d^2-\omega_c^2$ and the total frequency
$\omega_T=\sqrt{\omega_0^2+\omega_c^2}$, thus
\begin{equation}\label{hnws3}
\hat{H}_0+\hat{H}^\beta_{SO} =
\frac{\omega_0^2}{\omega_T^2}\frac{p_y^2}{2m^*}+\frac{p_x^2}{2m^*
}+\frac{m^* \omega_T^2}{2}(x-x_0)^2,
\end{equation}
where $x_0=s \frac{\omega_c p_y}{\omega_T^2 m^*}$, $s=\pm1$,
corresponds to the spin polarization along the $z$ direction. Hence
 we can conclude that 4-split
channels are present for a fixed Fermi energy, $\varepsilon_F$,
corresponding to $\pm p_y$ and $s_z=\pm1$.
 Notice the analogy with
the Hamiltonian corresponding to one electron in the QW when an
external transverse magnetic field is present.

\subsection{The Q1D ring}

Here we outline  briefly the derivation of the Hamiltonian
describing the motion of an electron in a realistic Q1D
ring\cite{meijer}. We consider the 2DEG in the $xy$ plane; then we
introduce a   radial potential $V_{\text{c}}(r)$, so that the
electrons  are  confined to move in a ring. The full
single-electron Hamiltonian reads
\begin{equation}
H= \frac{{\bf p}^{2}}{2m^*} +V_{\text{c}}(r)+H^\beta_{\text{so}},
\end{equation}
Due to the circular symmetry of the problem, it is natural to
rewrite the Hamiltonian in polar coordinates\cite{meijer}
\begin{eqnarray}
\nonumber
H&=&-\frac{\hbar^2}{2m^*}\left[\frac{\partial^2}{\partial r^2}+
\frac{1}{r}\frac{\partial}{\partial r}-\frac{1}{r^2}\left(i
\frac{\partial}{\partial \varphi} \right)^2
\right]+V_{\text{c}}(r)\\
\label{polarh} & & + \frac{\lambda^2}{\hbar}e\frac{E_r(r)}{r}\left(-i
\hbar \frac{\partial}{\partial \varphi} \right)\sigma_z,
\end{eqnarray}
because the electric field has just the radial component. It follows
that $L_z=-i \hbar \frac{\partial}{\partial \varphi}$  and
$\sigma_z$ commute with the Hamiltonian $\hat{H}$ and the
corresponding eigenvalues are $\pm \hbar \mu$ for $L_z$ and $\pm 1$
for $\sigma_z$.

 In the
case of a thin ring, i.e., when the radius $R_0$ of the ring is
much larger than the radial width of the wave function, it is
convenient to project the Hamiltonian on the eigenstates of
$$H_0=-\frac{\hbar^2}{2m^*}\left[\frac{\partial^2}
{\partial r^2}+\frac{1}{r}\frac{\partial}{\partial r}\right]+
V_{\text{c}}(r).$$ To be specific, we use a parabolic radial
confining potential
\begin{equation}
V_{\text{c}}(r)=\frac{1}{2} m^* {\omega_d^2}(r-R_0)^2 \quad ,
\end{equation}
for which the radial width of the wave function is given by
$l_\omega$. In the following, we assume $l_\omega/R_0 \ll 1$ and {
neglect contributions of order $l_\omega/R_0 $ to $H_0$} and to
the centrifugal term, $$H_c\simeq- \frac{ \hbar^2}{2 m^* R_0^2}\;
\frac{\partial^2}{\partial \varphi^2}=\hbar \omega_R
\frac{\partial^2}{\partial \varphi^2}.$$ In this limit, $H_0$
reduces to \bea \label{h0} H_0=-\frac{\hbar^2}{2m^*}
\left[\frac{\partial^2}{\partial r^2}\right]+\frac{1}{2} m^*
{\omega_d^2}(r-R_0)^2 \quad . \eea After some tedious calculations
(see appendix A) we are able to obtain the energy spectrum of
$H_0=H_c=H_{SO}$ as \bea\label{ene} \varepsilon_{n,\mu,s}\sim
\hbar \sqrt{\omega_d^2+2 \omega_c \omega_R \mu s} (n+\frac{1}{2})+
\hbar \omega_R \mu^2.\eea
\begin{figure}
\includegraphics*[width=.95\linewidth]{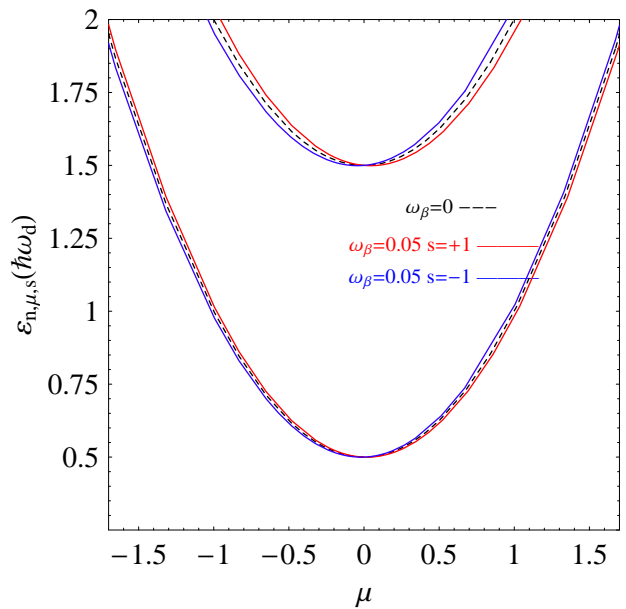}
 \caption {\label{fig1} Bandstructure with and without the effects of the SO coupling.
 Notice the splitting: for each value of the Fermi energy
$\varepsilon_F$ and for a fixed  band $n$, there are 4 different
eigenvalues. This spin dependent splitting in the energy allows
for the interference phenomena that we next discuss.}
\end{figure}
The corresponding bandstructure is shown in Fig.~\ref{fig1}. It
follows that for  fixed values of the Fermi energy,
$\varepsilon_F$, and of the band $n$ there are 4 different
eigenstates $ \Psi_{n,\mu}^s$ i.e. particles with fixed Fermi
energy $\varepsilon_{F}$ can go through the ring with four
different wave numbers $\pm \mu_{\pm 1/2}$, depending on spin and
direction of motion ($\pm $). Moreover  the presence of non
vanishing $\beta$ term implies an edge localization of the
currents depending on the electron spins, also giving the presence
of two localized  spin currents with opposite
chiralities\cite{noish}

 \

Now we want to remark the presence of a spin splitting which is the
basis of the interference phenomena in the transport through the
ring. In the typical Aharonov-Bohm devices the phase difference is
due to the enclosed  flux of an external magnetic field. In the
presence of a $\beta$ SO coupling the phase difference is
generated by the splitting of the opposite spin polarized
subbands.

 In
the presence of $\beta$ coupling the energy splitting is such that
particles with Fermi energy $\varepsilon_{F}$ can go through the
ring with four  different wave numbers $\lambda \mu_{\lambda,s}$,
depending on spin ($s$) and direction of motion ($\lambda=\pm $).
The quantities $\lambda \mu_{\lambda,s}$ are obtained by solving
$\varepsilon_{\mu,n}=\varepsilon_{F}$  and are not required to be
integer. Because  of the symmetry of the system, we can also obtain
that  $\mu_{+,\uparrow}=\mu_{-,\downarrow}$ and
$\mu_{-,\uparrow}=\mu_{+,\downarrow}$. Next the fundamental
quantity that we take in account is the phase difference $\pi
\Delta \mu $, where
$$
\Delta \mu= \mu_{+,\uparrow}-\mu_{-,\uparrow}=
\mu_{-,\downarrow}-\mu_{+,\downarrow}\sim 2 \frac{\omega_c}{
\omega_d}.
$$


\section{Theoretical approach to the transport through a ring}

\subsection{Ballistic transport and Landauer formula}

We first consider the case where the 1D ring of Sec.II.b is
symmetrically coupled to two contact leads (Fig.3.top panel left) in
order to study the transport properties of the system subject to a
constant, low bias voltage (linear regime). To this end, we
calculate the zero temperature conductance $G$ based on the Landauer
formula \cite{note9} \bea \label{G} G=\frac{e^2}{h} \sum_{n',n=0}^M
\sum_{\sigma',\sigma} T_{n' n}^{\sigma' \sigma} \; , \eea where
$T_{n' n}^{\sigma' \sigma}$ denotes the quantum probability of
transmission between incoming ($n,\sigma$) and outgoing
($n',\sigma'$) asymptotic states defined on semi-infinite ballistic
leads. The labels $n,n'$ and $\sigma,\sigma'$ refer to the
corresponding mode and spin quantum numbers, respectively. In our
case where $\sigma_z$ commutes with the Hamiltonian
$T^{\uparrow\downarrow}=T^{\downarrow\uparrow}=0$. We also limit our
analysis to the case of just one mode involved: $n=n'=0$.

\

The Landauer formula works in the ballistic transport regime, in
which scattering with impurities can be neglected and  the
dimensions of the sample are reduced below the mean free path of the
electrons. Here we think to ring conductors smaller than the
dephasing length $L_\phi$ i.e. with radius $R\lesssim 1 \mu m$ for
low temperatures ($T\ll1^oK$). We also assume that this regime is
not destroyed by the presence of just one obstacle as we will
discuss below. We want also point out that the Landauer formula in
the form of eq.(\ref{G}) works just at $T=0$ while a more general
formulation at finite temperatures has to take in account the width
of the distribution of injected electrons.

\subsection{Theoretical treatment of the scattering}

We approach this scattering problem using the quantum waveguide
theory\cite{2122}. For the strictly one dimensional ring the
wavefunctions in different regions for each value of the spin are
given below \bea \psi_I&=&e^{i k x_1}+r_s e^{-i k x_1} \nn \\
\psi_{II}&=&A_s e^{i \mu_{+,s} \varphi}+B_s e^{-i \mu_{-,s} \varphi} \nn \\
\psi_{III}&=&C_s e^{i \mu_{+,s} \varphi}+D_s e^{-i \mu_{-,s} \varphi} \nn\\
\psi_{IV}&=&t_s e^{i k x_2},  \nn \eea where we can assume the
wavevector of the incident propagating electrons in the leads
$k\sim \mu/R_0$.

Thus we use the Griffith boundary condition\cite{23}, which states
that the wave function is continuous and that the current density
is conserved at each intersection. Thus we obtain the transmission
coefficients.

\subsection{Interference and oscillations in the transmission}

Next we assume $\mu_{+,\uparrow}=\mu_{-,\downarrow}=\mu_0+\Delta
\mu$ and $\mu_{+,\uparrow}=\mu_{-,\downarrow}=\mu_0-\Delta \mu$
with $\Delta \mu$ depending on the strength of the $\beta$
coupling.

Thus we obtain the transmission coefficient as we show in Fig.2,
\begin{figure}
\includegraphics*[width=.95\linewidth]{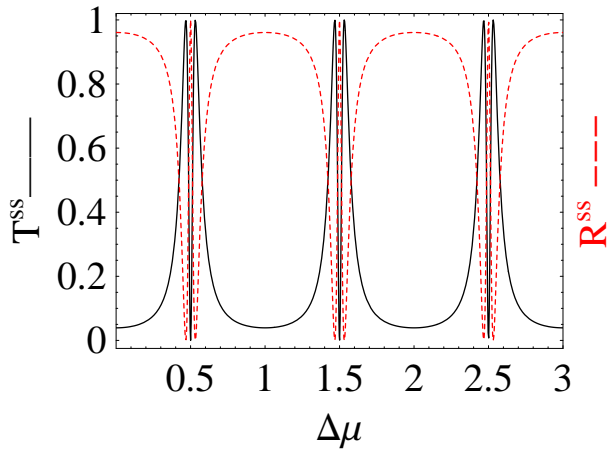}
 \caption {\label{fig3} Oscillations in the transmission (and reflection) due
 to the difference of the phases of the waves with opposite chiralities.
  This is a quite general result that does not depend on the cause which gives
  the phase difference, $\Delta \mu$.  }
\end{figure}
where the oscillations in the transmission are plotted as a
function of the difference of phase, rescaled by factor $\pi$, i.e. $\Delta \mu$.
Moreover it is clear that this kind of device is unable to produce
a spin polarized current, because it results
$T^{\uparrow\uparrow}=T^{\downarrow\downarrow}$.

\

The same result can be obtained by introducing a cut in the ring as
an infinite barrier at $\varphi=\varphi_B$. In Fig.3 we show the
oscillations in the transmission versus the position of the barrier.
\begin{figure}
\includegraphics*[width=.95\linewidth]{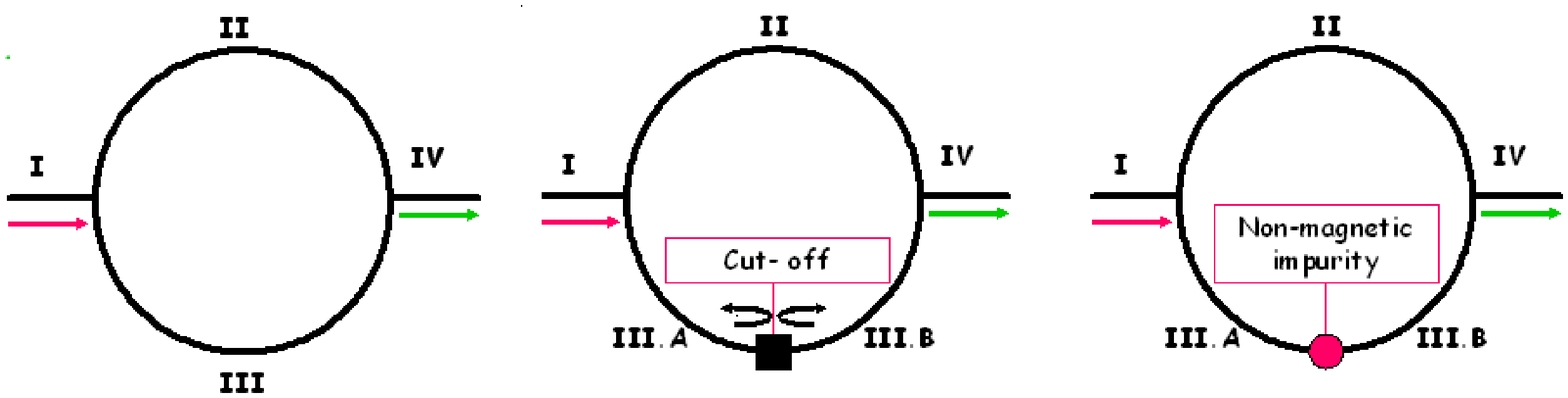}
\includegraphics*[width=.95\linewidth]{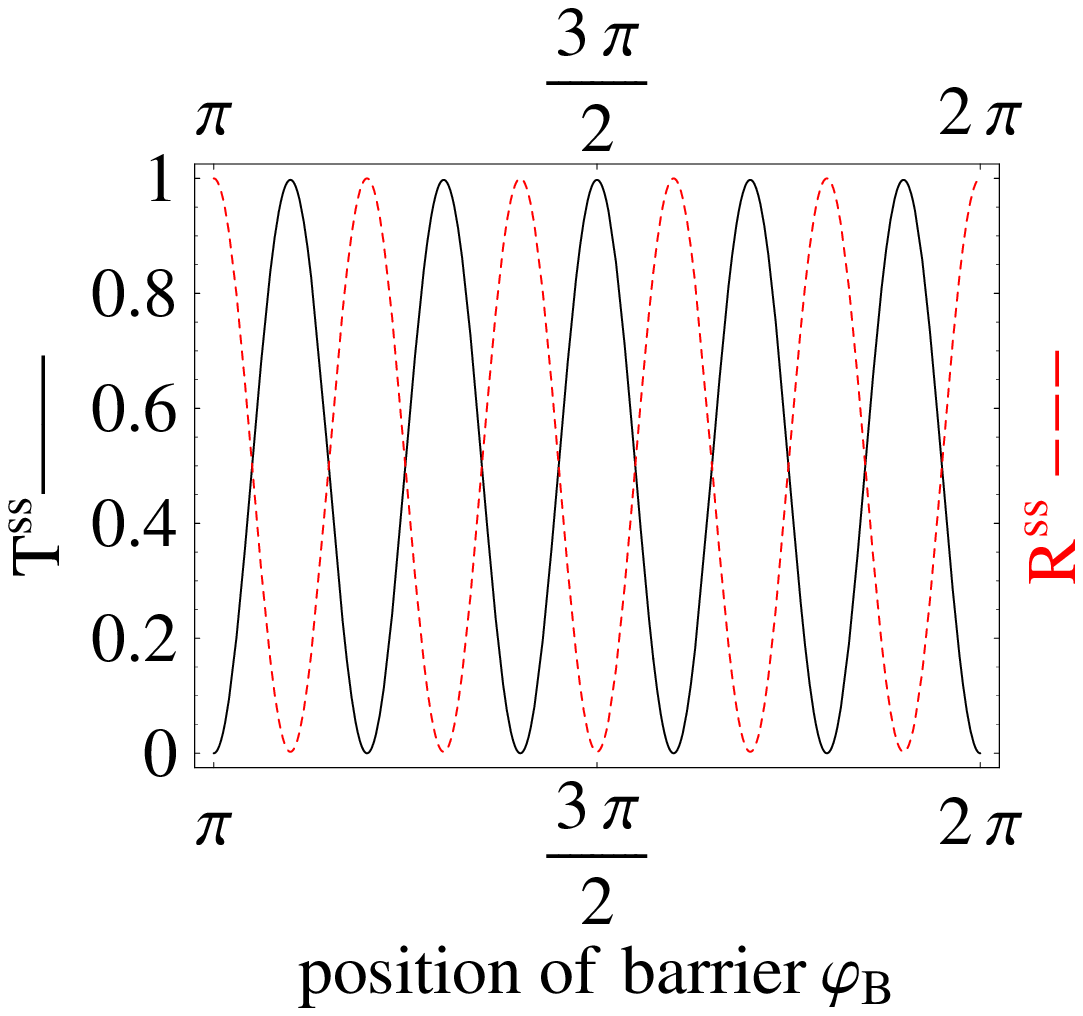}
 \caption {\label{fig3} (Top)(Left) Schematic diagram of a ring connected to two leads. (Center)
  Schematic draw of the ring with a cut off in $\varphi_B=3\pi/2$
 , i.e. interrupted by a totally reflecting barrier.
  On the right, the presence of a non magnetic obstacle.
  (Bottom) Oscillations in the transmission (and reflection) versus the cut position along the ring.}
\end{figure}
 Also in this case, there is no way to select a spin polarized
 current, because it results
$T^{\uparrow\uparrow}=T^{\downarrow\downarrow}$.

\section{Modulation of spin unpolarized currents}

Our main goal is to obtain a modulation of spin unpolarized
currents. In order to do that, we need a symmetry breaking for the
transport of  opposite spin polarized current, i.e.
$T^{\uparrow\uparrow}\neq T^{\downarrow\downarrow}$.

A central role, in order to pursue this goal, can be  played by the
presence of one or more obstacles along the path of the electrons
along the ring. This is the case of  impurities, disorder or
restrictions in the channel's width  (e.g. due to the presence of a
Quantum Point Contact along the channel). Next we analyze the
presence of just one obstacle and we name it {\it single non
magnetic obstacle} in analogy to the non magnetic impurity discussed
in ref.[\onlinecite{SGZ03}].

\subsection{Effects of a non-magnetic obstacle}

In order to discuss the effect of a non-magnetic obstacle on the
transmission of the ring, we have to introduce a correction in our
model. To simplify the problem, we will now assume that the obstacle
is a delta-function barrier $V_0 \delta(\varphi-\varphi_B)$. Thus we
can calculate the transmission by imposing the boundary conditions.
Results are reported in Fig.4.
\begin{figure}
\includegraphics*[width=.95\linewidth]{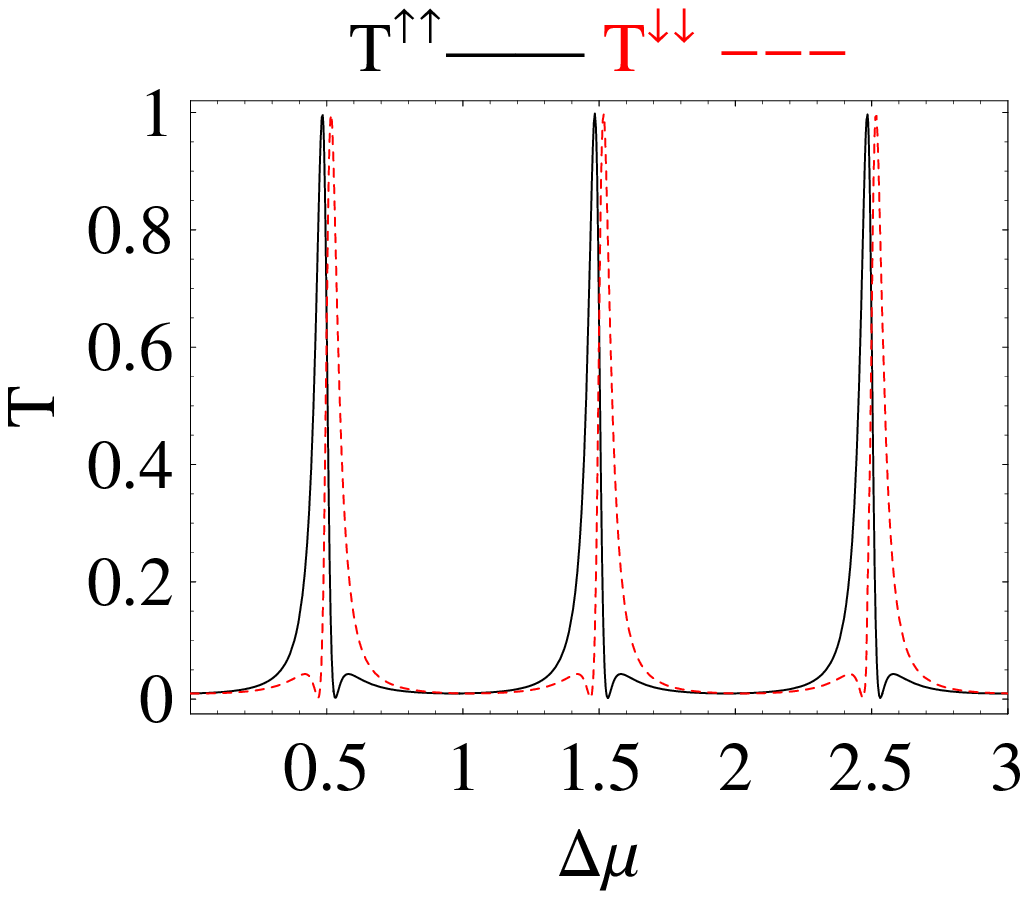}
\includegraphics*[width=.95\linewidth]{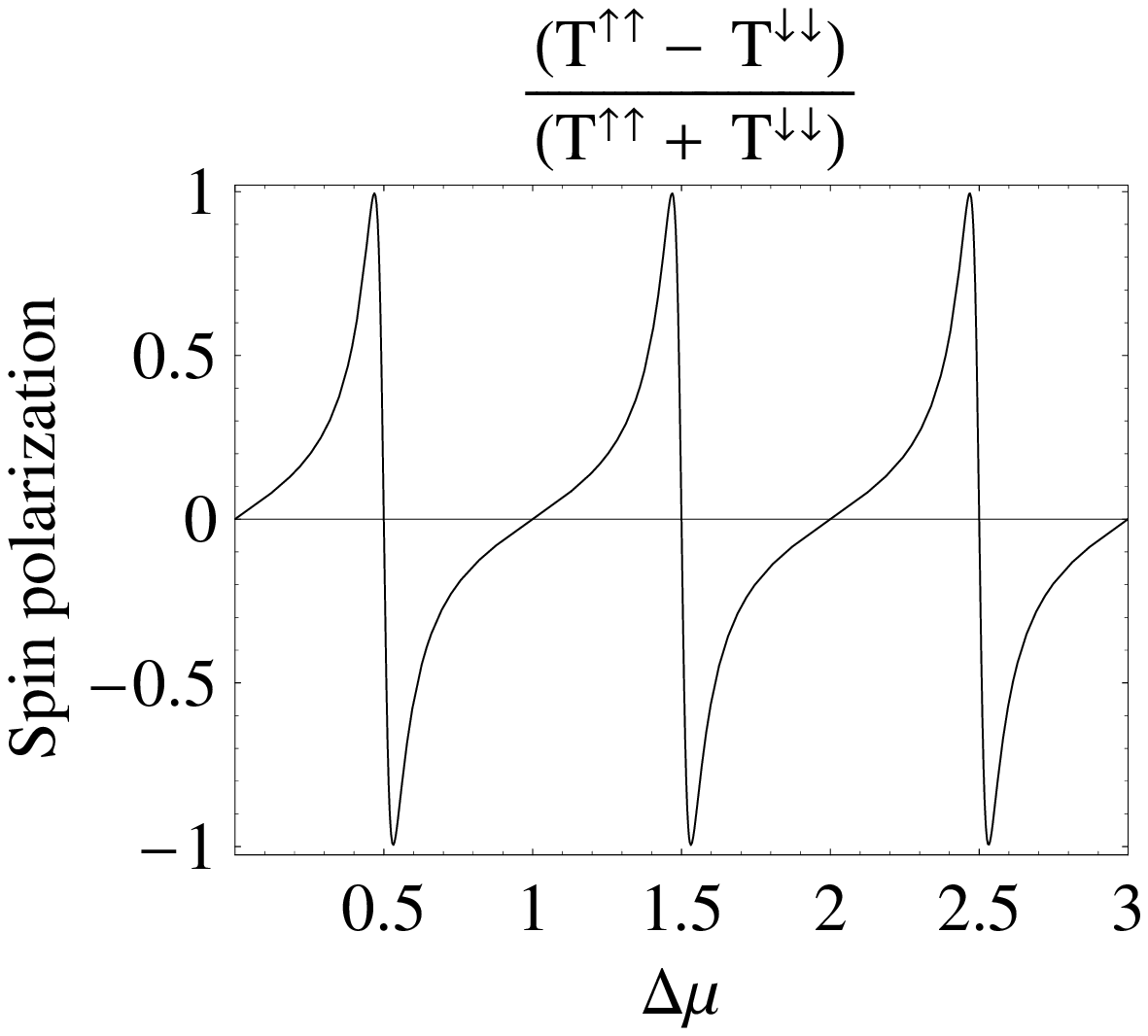}
 \caption {\label{fig3} Transmission of the ring with a non magnetic obstacle at $\varphi_B=3\pi/2$.
 (Top) The spin  transmissions versus the phase difference $\Delta \mu$.
 (Bottom) The extrapolated polarization, $P_z$ given in Eq. (\ref{pz}) of the emerging current.}
\end{figure}
In the presence of the obstacle the symmetry between the opposite
spin polarization is broken and the transmission
$T^{\uparrow\uparrow}$ differs from $T^{\downarrow\downarrow}$ (see
Fig4.top). It follows that a spin polarized current can be observed
at any values of $\Delta \mu$. Thus  in the presence of just one
obstacle the ring is able to select a polarized current.

\

From Fig.4 it is evident that the transmission polarized spin
current is controlled by the phase difference, as well as by the
modulation, analogously to the transmission charge current. In
order to see this modulation clearly, we introduce a dimensionless
quantity $P_z$ to describe the polarization along the $S_z$ spin
axis of current transmitted through the Q1D ring, which is defined
by \bea \label{pz} P_z=P_z(\pi \Delta
\mu)=\frac{j_\uparrow-j_\downarrow}{j_\uparrow+j_\downarrow}=
\frac{T^{\uparrow\uparrow}-T^{\downarrow\downarrow}}{T^{\uparrow\uparrow}+
T^{\downarrow\downarrow}}.\eea Here the spin resolved currents
$j_s$ were obtained employing the Landauer formula and  $P_z$
turns out to be independent from the momentum of the incident
charge carriers. This could yield an important advantage for
device applications. Here $P_z$ is similar to the spin injection
rate defined in ferromagnetic/semiconductor/ferromagnetic
heterostructures\cite{233}, and it can be measured
experimentally\cite{apl}.

\begin{figure}
\includegraphics*[width=.95\linewidth]{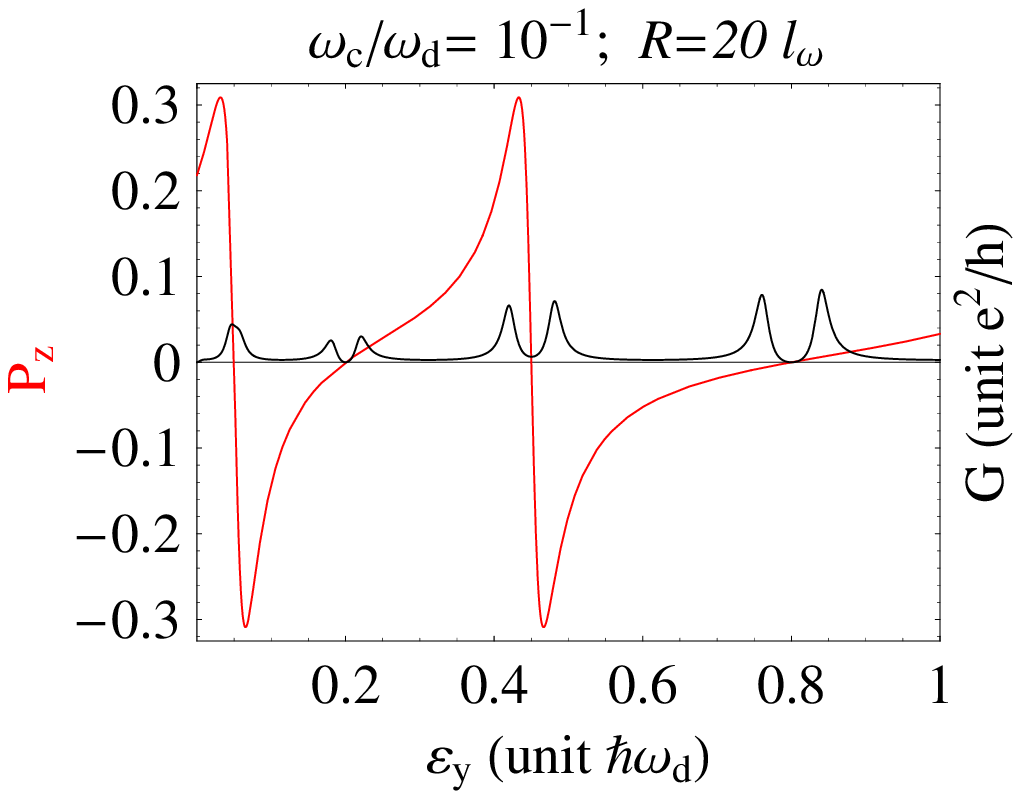}
\includegraphics*[width=.95\linewidth]{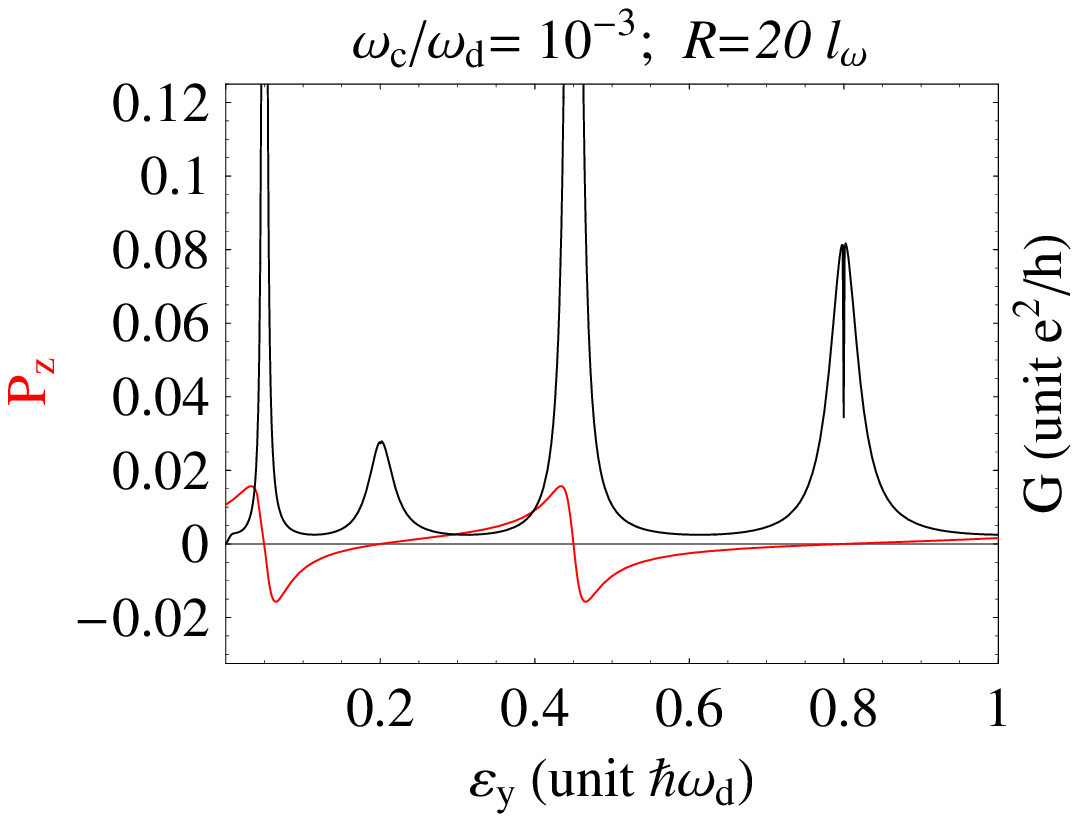}
 \caption {\label{fig7} The spin polarization $P_z$ of the current exiting from a 1D ring (red
 line), an the charge conductance,
  as functions of the Fermi energy $\varepsilon_F$  for two realistic strength of the SO-$\beta$ coupling.
We can observe that the presence of peaks in the spin polarization
is related to dips in the  charge transport. These dips in $G$
should correspond to the values of the Fermi energy which give
integer angular momenta ($\mu=n$), but the presence of spin
splitting doubles the peaks because of the symmetry breaking. We
observe that the maximum spin polarization is obtained near the odd
peaks.}
\end{figure}

\section{Modulation of spin current based on the Rashba SO coupling}

\subsection{$\alpha$-SO coupling}
In what follows we take in account  the {\it natural} $\alpha$
(Rashba) coupling~\cite{Kelly}. In semiconductors
heterostructures, where a 2DEG is confined in a potential well
along the $z$ direction, the SO interaction is of the type
proposed by Rashba \cite{rashba}: it arises from the asymmetry of
the confining potential which occurs in the physical realization
of the 2DEG, e.g.  due to the band offset between $AlGaAs$ and
$GaAs$. In this case the SO Hamiltonian in Eq.\ref{H_SO} becomes
\begin{equation}
\label{hal}
H_{\alpha}=\frac{\alpha}{\hbar}\left(\sigma_{x}\,p_{y}-\sigma_{y}\,p_{x}\right),
\end{equation}
that in polar coordinates can be written as
\begin{equation}
\label{hal2} H_{\alpha}=-\frac{\alpha}{r} \sigma_r \left( i
\frac{\partial}{\partial{\varphi}} \right)+i \alpha \sigma_\varphi
\frac{\partial}{\partial r}-\frac{i}{2}\frac{\alpha}{r}
\sigma_\varphi.
\end{equation}
Here $\sigma_r=\cos\varphi\,\sigma_x+\sin\varphi\, \sigma_y$ and
$\sigma_\varphi=-\sin\varphi\,\sigma_x+\cos\varphi\, \sigma_y$. In
the case of a thin ring, i.e. in the strictly one-dimensional
case, when the radius $R$ of the ring is much larger than the
radial width of the wave function $\l_\omega$, we can neglect the
second term in the r.h.s. of Eq.\ref{hal2} and assume $r=R$, in
agreement with the result in Eq. (2) of ref.(\onlinecite{FRURIC}).

As in the case of $\beta$ coupling we can introduce an effective
magnetic field which in this case is oriented in the plane of the
ring.

\subsection{Energy bands and wavefunctions}

After some tedious calculations  we are able to obtain the energy
spectrum\cite{FRURIC} as \bea\label{ene2} \varepsilon_{n,\mu,s}&=& \hbar \omega_d
(n+\frac{1}{2})+  \hbar \omega_R
\left(\mu+\frac{1}{2}\right)^2+\frac{\hbar \omega_R}{4}\nn \\ & +&s
{\hbar} \sqrt{\omega_R^2+\omega_\alpha^2 }|\mu+\frac{1}{2}| ,\eea
where $\omega_\alpha=\alpha/(\hbar R)$ and $s$ is the spin
polarization.
If we introduce  $j\equiv \mu+1/2$ eq.(\ref{ene2}) becomes
\bea\label{ene3} \varepsilon_{0,j,s}= \frac{\hbar \omega_d}{2}+
\hbar \omega_R \left(j-s \frac{j_\alpha}{2}\right)^2+\hbar
\frac{\omega_\alpha^2}{\omega_R},\eea where $j_\alpha\equiv
\sqrt{1+\frac{\omega_\alpha^2}{\omega_R^2}}$.

 It follows that for a fixed value of  the Fermi energy,
$\varepsilon_F$, there are 4 different eigenstates $ \Psi_{\pm,0,\mu}^s$
i.e. particles  can go
through the ring with four different wave numbers $\mu^s_{\pm,s}$,
depending on spin ($s$) and direction of motion ($\pm $) as in the
case discussed in the previous sections.
The wave numbers can be obtained by solving the equation
$$
\tilde{\varepsilon}=\hbar \omega_R(\mu +
\frac{\Phi_{AC}^\pm}{2\pi})^2,
$$
where $\tilde{\varepsilon}\equiv\varepsilon_F-\hbar
\omega_d/2-?\hbar{\omega_\alpha^2}/{\omega_R}$ and
$\Phi_{AC}^\pm=-\pi(1\pm J_\alpha)$ are the Aharonov Casher phase
which are acquired while the two spin states evolve in the ring in
the presence of the Rashba electric field.

 The main difference with the $\beta$ case  is
that the spin are now polarized in a different direction i.e.
$\hat{\bf s}_\alpha$ with an angle $2\theta$ respect to the $z$ axis
corresponding to $\tan(2\theta)= \frac{\omega_\alpha}{\omega_R}$.

Thus that we can write the wavefunctions as \bea \Psi_{\pm,0,\mu}^+
&=&u_0(r)e^{i \mu_{\pm,+} \varphi}\left(\begin{array}{c}
 \cos(\theta) \\
 \sin(\theta)e^{i\varphi}
\end{array}\right) \;\;\; \nn \\
\Psi_{\pm,0,\mu}^- &=&u_0(r)e^{i
\mu_{\pm,-}\varphi}\left(\begin{array}{c}
 \sin(\theta) \\
 -\cos(\theta)e^{i
\varphi}
\end{array}\right) \;\;\; \nn
\eea

Thus  fundamental quantity which gives  the phase difference is $
\Delta \mu \pi $ is now given by $ \Delta \mu= j_\alpha-1$.
 \
\subsection{From the transmission to the conductance }

Now we can develop the calculations based on the Landauer formula
in order to obtain the zero temperature conductance as discussed
in section III. This approach, as we discussed above, is based on
the calculation of the transmission amplitudes
$T^{ss'}=|t^{ss'}|^2$. Thus we have to solve the scattering
problem analogously to the case reported in section III by using
the quantum waveguide theory.

Next we assume that the spin polarization along $\hat{\bf
s}_\alpha$ is a constant of motion, thus $t^{+-}=t^{-+}=0$. Now we
can write the coefficients $t$ in the two different basis as \bea
t^{\uparrow\uparrow} &= &
\cos^2({\theta}) t^{++} +  \sin^2({\theta})t^{--} \nn \\
t^{\uparrow\downarrow} &= &
-\cos({\theta})\sin({\theta})t^{++}+\cos({\theta})\sin({\theta})t^{--}
\nn \\
t^{\downarrow\downarrow}&= &
-\sin^2({\theta})t^{++}-\cos^2({\theta})t^{--}
\nn \\
t^{\downarrow\uparrow}&= &
\cos({\theta})\sin({\theta})t^{++}-\cos({\theta})\sin({\theta})t^{--}.
 \label{matt} \eea
It follows that $t^{\uparrow\downarrow}=-t^{\downarrow\uparrow}$.

 \

 \subsection{Modulation of a spin current}

 Our main goal is to obtain a modulation of spin unpolarized
currents. In order to do that, we need a symmetry breaking for the
transport of  opposite spin polarized current, i.e.
$T^{\uparrow\uparrow}\neq T^{\downarrow\downarrow}$ or
$T^{\uparrow\downarrow}\neq T^{\uparrow\downarrow}$. The equations
in Eq.(\ref{matt}) showed that if $t^{++}=t^{--}$ there is no
symmetry breaking, in fact also
$t^{\uparrow\downarrow}=-t^{\downarrow\uparrow}$. Thus as in the
case of the $\beta$ coupling no spin polarization is present when
we consider a clean ring.

\

A central role, in order to obtain a modulation of spin unpolarized
currents, can be  played by the presence of one or more obstacles
along the path of the electrons in  the ring as we discussed above.
The corresponding symmetry breaking gives a significant spin
polarization of the transmitted current
$$
P_z=\frac{T^{\uparrow\uparrow}-T^{\downarrow\downarrow}}{T^{\uparrow\uparrow}+T^{\downarrow\downarrow}+2
T^{\uparrow\downarrow} }.
$$
It follows that a spin polarized current can be observed due to the
Rashba phase shift (see Fig.7). Thus in the presence of just one
obstacle the ring is able to select a polarized current. However by
a comparison with the plots corresponding to the $\beta$ coupling it
seems clear that a $\beta$-coupling based mechanism could be  more
efficient in obtaining a spin polarized current.

\begin{figure}
\includegraphics*[width=.95\linewidth]{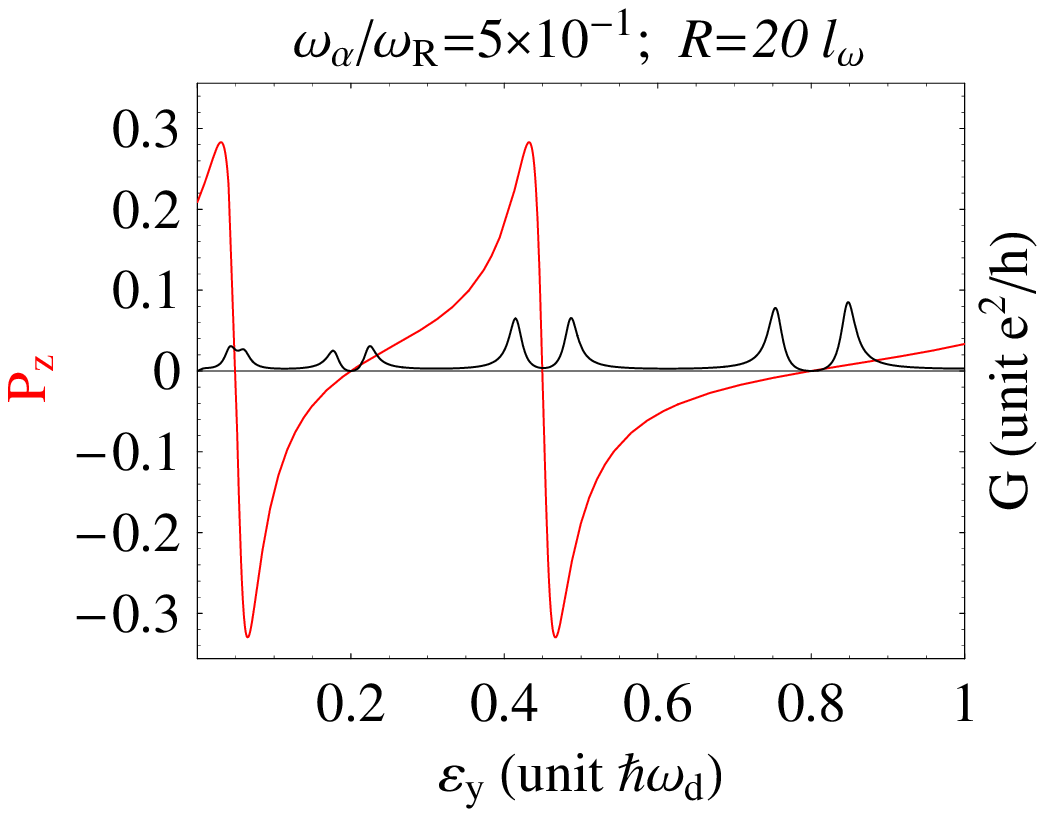}
\includegraphics*[width=.95\linewidth]{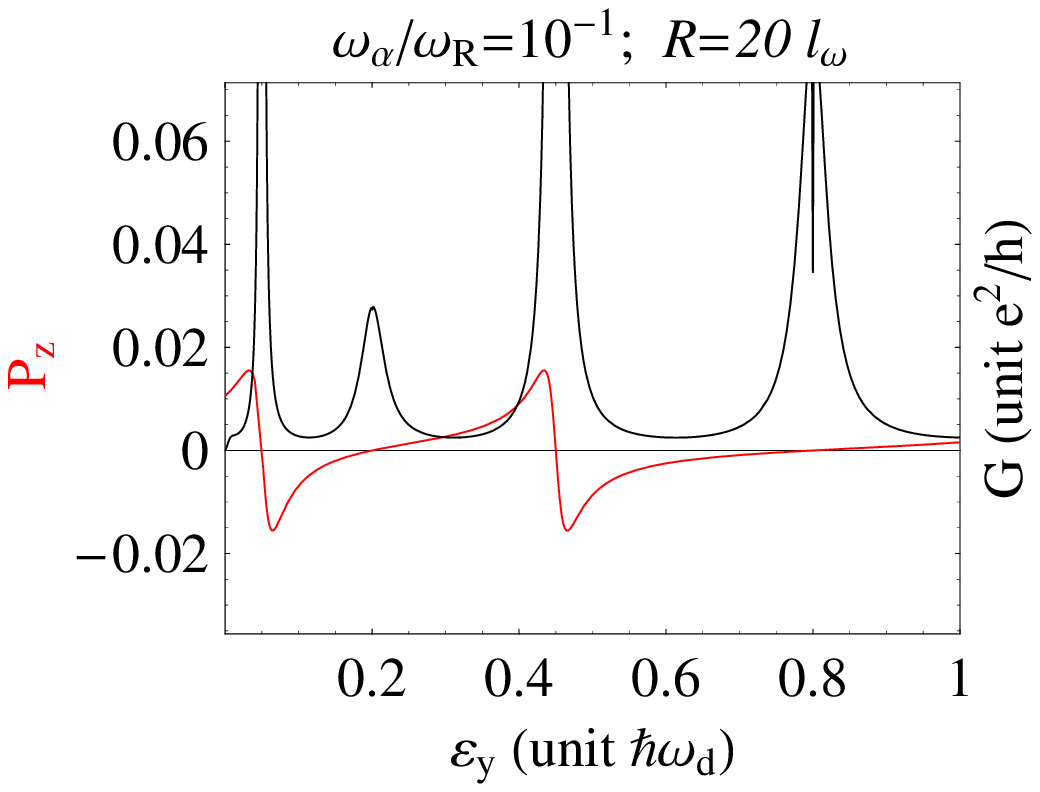}
 \caption {\label{fig7} The spin polarization $P_z$ of the current exiting from a 1D ring (red
 line), an the charge conductance,
  as functions of the Fermi energy $\varepsilon_F$  for two realistic strength of the SO-$\alpha$ coupling.
We can observe that the presence of peaks in the spin polarization
is related to dips in the  charge transport. These dips in $G$
should correspond to the values of the Fermi energy which give
integer angular momenta ($\mu=n$), but the presence of spin
splitting doubles some of the peaks, because of the symmetry
breaking. However this splitting is clear just for strong values
of the coupling. We observe that the spin polarization is
significant near the odd peaks, whereas it vanishes in
correspondence of the even peaks.}
\end{figure}

\section{Discussion}

The ring conductors have played an essential role in observing how
coherent superpositions of quantum states (i.e.,
quantum-interference effects) on the mesoscopic scale leave imprints
on measurable transport properties\cite{[26]}. In fact they
represent a solid state realization of a two-slit experiment, where
an electron entering the ring can propagate in two possible
directions (clockwise and counterclockwise). In these devices
superpositions of  quantum states are sensitive to the acquired
topological phases in a magnetic [Aharonov-Bohm effect] or an
electric [Aharonov-Casher  effect for particles with spin] external
field whose variations generate an oscillatory pattern of the ring
conductance \cite{[27]}.

\

In this paper we found that a non vanishing spin polarized current
can be measured for a 2 leads ballistic ring  in the presence of the
{\it natural} $\alpha$ and  $\beta$ term of the SO coupling. As we
showed in Figs.5 and 6, some peaks in the spin polarization,
$P_{z}$, are present near the measurable peaks in the  charge
conductance.All of our calculations are limited to the lowest
subband but can be easily extended to the several subband case.

\begin{figure}
\includegraphics*[width=.95\linewidth]{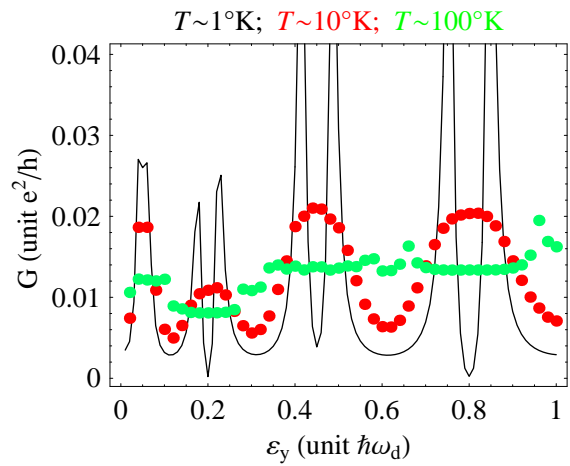}
 \caption {\label{fig8} The charge conductance for three different scales of temperatures between
 $0°K$ and some hundred degrees. }
\end{figure}
Moreover, in order to observe these oscillations at finite
temperatures, the width of the distribution of injected electrons
should not exceed the gap between the adjacent peaks of $G$  in
Figs.5 and 6, while its center (i.e., $\varepsilon_F$ of the
reservoirs) should be adjusted to their position\cite{nprl}.
However for the spin-filter realization  it is relevant to
evaluate the efficiency of the device at non-zero temperature.
Thus in the following we generalize our calculations at finite
temperature $T$. The conductance at finite $T$ is given
by\cite{citro}
\begin{equation}
G=-(e^2/h)\sum_{\sigma} \int_0^{\infty} d\varepsilon \frac{\partial
f(\varepsilon,\varepsilon_F,T)}{\partial
\varepsilon}|T_\sigma(\varepsilon_F)|^2,
\end{equation}
where $f$ is the Fermi distribution function and $T$ the
temperature. As we show in Fig.7 the peaks disappear when the
temperature becomes larger than some tens of $°K$. Thus the proposed
mechanism for the spin polarization works just at low temperatures.

\

In several papers (e.g. refs.(\onlinecite{nprl,FRURIC})) it was
discussed how the tuning of the Rashba SO coupling in a
semiconductor heterostructure hosting the ring generates
quasiperiodic oscillations of the predicted spin-Hall current, due
to spin-sensitive quantum-interference effects caused by the
difference in the Aharonov-Casher phase accumulated by opposite
spin states. In those cases an additional external field was
needed in addition to the natural Rashba coupling. The authors of
Refs.(\onlinecite{nprl,FRURIC})) proposed that the  value of the
$\alpha$ SOC could  be tuned by controlling the transverse
electric field by giving   $\omega_\alpha/\omega_R$  in the range
$0-10$. In the present work we discussed  the transport in the
presence of one non-magnetic obstacle in the ring with just the
natural  SO couplings, where the spin polarization of the current
is governed by the gate voltage modulation. We demonstrated that a
 spin polarized current can be induced when an unpolarized charge current
is injected in the ring thanks to the presence of the obstacle.

\

In section II and IV of this paper we assumed the $\alpha$-coupling
to be negligible, although in general this term is comparable to (or
larger than) the $\beta$-coupling term.
 By comparison with typical quantum-well and transverse
electric fields, the SO-coupling constant $\beta$ can be roughly
estimated as at least $\beta \sim 0.1\,\alpha$\cite{morozb}.
Moreover, in square quantum wells where the value of $\alpha$ is
considerably diminished~\cite{sw}, the constant $\beta$ may well
compete with $\alpha$. Furthermore the effects of the Rashba term on
the spin polarization are often significant just for strong values
of $\alpha$, some order of greatness  larger than $\sim 10^{-11}\;
eV\; m$("natural" values of $\alpha$ at  the GaAs
interface\cite{Nitta}) while the in plane $\beta$ coupling gives a
good spin polarization in the currents also for small values of
$\beta$, that are however larger than the usual ones (see
ref.[\onlinecite{noiq}]).

It is clearly more difficult to modulate the strength of the
$\beta$-SO coupling by acting on the split gate voltage. Thus the
feasibility of a $\beta$ governed device  mainly depends on its size
and on the materials. The fundamental theoretical parameter in
section IV, $\Delta\mu$, is  proportional to the the ratio
$\omega_c/\omega_d$, corresponding to $\lambda^2/l_\omega^2$, i.e.
the ratio  between a material dependent parameter $\lambda$ and a
size  dependent one $\l_\omega$ (that can be assumed to be a
fraction of the  real width,$W$, of the conducting channel). The SO
strengths have been theoretically evaluated for some semiconductors
compounds. In a QW ($W\sim 100$) patterned in InGaAs/InP
heterostructures, where $\lambda^2$ takes values between $0.5$ and
$1.5\,nm^2$, it results $\hbar \omega_{c}\sim 10^{-6}-10^{-4} eV$,
corresponding to $\omega_{c}/\omega\sim 10^{-4}-10^{-3}$ as in InSb,
where $\lambda^2\sim 500 {\AA}^2$.
 For GaAs heterostructures, $\lambda^2$ is one order
of magnitude smaller ($\sim 4.4 {\AA}^2$) than in InGaAs/InP,
whereas for HgTe based heterostructures it can be more than three
times larger\cite{HgTe}. However, the lithographical width of a
wire defined in a 2DEG can be as small as $20nm$\cite{kunze}; thus
we can realistically assume that $\omega_{c}/\omega_d$ runs from
$1 \times 10^{-6}$ to $1\times 10^{-1}$[\onlinecite{notaf}]. Here
we can realistically assume that the ring has a width of just some
tens of $nm$s.

\

The case reported in section V is more simple to be realized because
in typical materials natural  $\alpha$ is larger than $\beta$ and
can also be tuned by controlling the transverse electric field. The
phase shift is proportional to $\omega_\alpha/\omega_R$ so that a
further modulation of the phase shift  can be obtained by acting on
the ring's radius.

\

Thus, we can propose the discussed  devices as  spin filters based
on the Q1D ring. We showed how the spin filtering is grounded on the
presence of a non magnetic obstacle which produces a more or less
spin polarized current.
 However, also in
samples where spin polarization  is quite smaller, the efficiency of
a two leads ring as a spin filter can be amplified by realizing a
series of these devices.

\

We acknowledge the support of the grant 2006 PRIN "Sistemi
Quantistici Macroscopici-Aspetti Fondamentali ed Applicazioni di
strutture Josephson Non Convenzionali".

\appendix
\section{Spectrum and spin splitting }
Next we can introduce the new variable $\xi=r-R_0$ ($p_\xi=-i \hbar
\frac{\partial}{\partial r}$). The eigenvalues of $L_z$ are $\hbar
\mu$ and the ones of $\sigma_z$ are $s$ ($S_z\equiv
\frac{\hbar}{2}\sigma_z$). Thus we can write \bea
H_0+H_c+H^\beta_{SO}&\simeq&\frac{p_\xi^2}{2
m^*}+\frac{m^*}{2}\omega_d^2 \xi^2+\frac{m^*}{2}\omega_\beta^2 s \mu
\xi^2 \nn \\& & -\frac{m^*}{2}\omega_\beta^2 s \mu 2 R_0 \xi+ \hbar
\omega_R \mu^2 ,\eea where $\omega_\beta^2\equiv \frac{\beta
}{m^*l_\omega R_0^2}$. Now we can introduce the new variables
$\omega_T(\mu,s)^2\equiv\omega_d^2+\omega_\beta^2 s \mu$ and
$\xi_0(\mu,s)=\mu s\frac{\omega_\beta^2}{\omega_T(\mu,s)^2}R_0$, in
order to obtain \bea H &\simeq& \frac{p_\xi^2}{2
m^*}+\frac{m^*\omega_T(\mu,s)^2}{2}\left(\xi-\xi_0(\mu,s)\right)^2
+ \hbar \omega_R \mu^2 \nn \\
& &-\frac{m^*\omega_\beta^4R_0^2}{2\omega_T(\mu,s)^2}, \nonumber
\eea from which the energy spectrum follows,  \bea\label{ene}
\varepsilon_{n,\mu,s}= \hbar \omega_T(\mu,s)(n+\frac{1}{2})+ \hbar
\omega_R \mu^2 -\frac{m^*\omega_\beta^4
R_0^2}{2\omega_T(\mu,s)^2}.\eea

It follows that for  fixed values of the Fermi energy,
$\varepsilon_F$, and of the band $n$ there are 4 different
eigenstates which have the general form
$$
\Psi_{n,\mu}^s=u_n\left(r-R_0-\xi_0(s,\mu)\right)e^{i \mu \varphi
}\chi_s,
$$
where $u_n(x)$ are the eigenstates of the 1D harmonic oscillator.

As we showed in ref.\onlinecite{noish} the presence of non vanishing
$\beta$ term implies an edge localization of the currents depending
on the electron spins, also giving the presence of two localized
spin currents with opposite chiralities. However, in our
calculations
 we assume $u_n\left(\xi-\xi_0(s,\mu)\right)\simeq
u_n\left(\xi+\xi_0(s,\mu)\right)$,  in order to reduce the problem
to a strictly one dimensional one.

\bibliographystyle{prsty} 

\bibliography{}

\begin{thebibliography}{99}
\bibitem{notaf} In any case $W$ should be larger than $\lambda_F$, so that
at least one conduction mode is occupied.
\end{thebibliography}

\end{document}